\documentclass[preprint2]{aastex}
\slugcomment{To appear in Astrophys. Journal}

\shorttitle{H$\alpha$ Velocity Mapping of the Stephan's Quintet}
\shortauthors{C. M. Gutierrez et al.}

\begin{document}

\title{New light and shadows on Stephan's Quintet}

\author{C. M. Guti\'errez}
\affil{Instituto de Astrof\'{\i}sica de Canarias, E-38205 La Laguna, Tenerife, Spain}
\email{cgc@ll.iac.es}
\author{M. L\'opez-Corredoira}
\affil{Instituto de Astrof\'{\i}sica de Canarias, E-38205 La Laguna, Tenerife, 
Spain\\ 
Astronomisches Institut der Universit\"at Basel, Venusstrasse 7, CH-4102
Binningen, Switzerland}
\author{F. Prada}
\affil{Centro Astron\'omico Hispano-Alem\'an, Apdo. 511, E-04080 Almer\'\i
a, Spain}
\and
\author{M. C. Eliche}
\affil{Instituto de Astrof\'{\i}sica de Canarias, E-38205 La Laguna, Tenerife, Spain}

\begin{abstract}

We present deep broad-band $R$ and narrow-band H$\alpha$ images of
Stephan's Quintet. The observations in the $R$ band show that the
diffuse halo of the Stephan's Quintet is larger than previously thought
and extends out to NGC~7320C. However, we have not found emission
connecting NGC~7331 and NGC~7320 to $R\sim 26.7 $ mag/arcsec$^2$ (at
more than 3-$\sigma$ level), so there is no direct evidence up to this
limiting magnitude of a relation between the peculiar kinematical
structure found in NGC~7331 and an ongoing or past interaction between
this galaxy and NGC~7320. The H$\alpha$ emission at high-velocity
(6000--7000 km s$^{-1}$) is distributed in a diffuse structure running
NS between NGC~7319 and NGC~7318B and in some other more concentrated
features. Some of these are located in the tidal tails produced by the
interaction between the galaxies of the group.  With the H$\alpha$
images we have made a two-dimensional velocity map which helps to
identify the origin of each structure detected. This map does not show
features at intermediate velocities between the high- and low-redshift
members of the group.  This is in agreement with the standard scenario
in which the apparent proximity of NGC~7320  to the rest of the
galaxies of the Quintet is merely a projection effect. The only point
which is unclear in this interpretation, is an H$\alpha$ filament which
is seen extending throughout NGC 7320 with velocity at 6500 km/s
instead of the 800 km/s expected for this galaxy.

\end{abstract}



\section{Introduction}

Stephan's Quintet (SQ) was discovered in 1877 and is one of the most
popular and studied groups of galaxies. It is classified as a compact
group (HCG 92) in the Hickson (1982) catalog. The core of the group is
formed by three galaxies (NGC~7319, NGC~7318A, and NGC~7317), which
have recessional velocities $\sim$ 6600 km s$^{-1}$ with very low
velocity dispersions. The other members of the group are NGC~7318B, at
$\sim$ 5700 km s$^{-1}$, and NGC~7320  at $\sim$800 km s$^{-1}$.
According to the orthodox view, NGC~7320 is a foreground galaxy, while
NGC~7318B which has a difference in velocity of $\sim$1000 km s$^{-1}$
with respect to the core of the group, shows clear signs of strong
distortions which indicate ongoing interaction with the group. Two
other galaxies NGC~7320C ($\sim$ 6000 km s$^{-1}$) and NGC~7331 ($\sim$
800 km s$^{-1}$) are not considered in the Hickson scheme as members of
the group, but their apparent proximity and several reasons that we
will discuss in this  paper make it convenient to consider them here. A
diagram showing the relative positions of the galaxies in the SQ and
NGC~7331 can be found in Moles, Sulentic, \& Marquez (1997).

In spite of the large observational effort dedicated to the study of SQ
from the radio  to the X ray continuum, there are many points which yet
require further attention.  In particular, in this paper we address the
following specific questions: {\it i\/}) whether  the apparent
proximity of NGC~7320 to the other members of  SQ is merely a
projection effect; {\it ii\/}) whether there is any evidence of
interaction between  SQ (specifically NGC~7320) and NGC~7331 that could
explain the particular kinematics of this galaxy (Prada et al.  1996);
and {\it iii\/}) whether  the core of  SQ (NGC~7318A, NGC~7318B and
NGC~7319) is connected with NGC~7320C. We have tried to answer these
questions by conducting an observational program of broad- and
narrow-band imaging. The narrow-band H$\alpha$  images analyzed in this
paper cover velocities in the range 0--8000 km s$^{-1}$, while the
broad-band $R$ observations correspond to two different regions: the
south part of NGC7331 and SQ respectively, covering a total area
$\sim$650 square arcminutes. With the $R$ images we have investigated
the possible presence of bridges between NGC~7331 and SQ, and with the
narrow-band observations  we have mapped the H$\alpha$ and derived a
map with the velocity field of this emission in SQ.

Section~2 of this paper summarizes our present knowledge and previous
observations of  SQ.  Section~3 presents the observations and data
processing. In Sections~4 and 5 we analyze the broad- and narrow-band
images, respectively. Section~6 discusses our results in the context of
earlier observations. Our conclusions are given in Section 7.

\section{Present status of observations of Stephan's Quintet}

\subsection{\small Optical and near IR}

Recent studies have been conducted in the optical and near infrared
from ground-based telescopes and from {\it ISO} and the {\it Hubble Space
Telescope} ({\it HST\/}). From {\it HST} broad-band images, Gallagher 
et al. (2001) have recently claimed the discovery of a large number of
young star clusters at distances up to 20 kpc from the  galactic center
in the tidal debris of NGC~7319. This is in agreement with the standard
dynamical picture (Moles, Marquez, \& Sulentic 1998), in which the ongoing
interaction with NGC~7319 is older than that occurring now between
NGC~7318A and NGC~7318B. 

Ohyama et al. (1998) have discovered two emission-line regions in the
tidal arm of NGC~7318B in a position which coincides with previously
discovered radio continuum emission. The lines in these two region are
at 6560 and 6720 km s$^{-1}$, respectively, and much wider (900 km
s$^{-1}$) than the normal H~{\scshape ii} regions in NGC 7318B and NGC
7319. From this, and from the the ratio between the S~{\scshape ii} and
the H$\alpha$ emission, they conclude that the emission is the remnant
of a large number of recent supernovae. Plana et al. (1999) have
presented H$\alpha$ observations in  SQ with a Fabry--Perot
interferometer. From these, they have detected and derived the
velocities for 23 H$\alpha$ regions with velocities in the range
5540--6700 km s$^{-1}$. Mendes de Oliveira et al. (2001) and
Iglesias-P\'aramo \& V\'\i lchez (2001) have recently presented
evidence that some of this emission corresponds to dwarf galaxies
located within tidal tails in the SQ.

%Molecular gas

Yun et al. (1997) have demonstrated that SQ is deficient in molecular gas.  CO emission is detected only in NGC 7319 and is concentrated mostly in
a region 8 kpc north of the center of the galaxy. This asymmetry could
be produced by recent tidal disruptions. However Gao \& Xu (2000),
using BIMA, have detected a larger abundance of CO in NGC 7319, and
also in NGC 7318B and the intergroup starburst region. Finally, Smith \& Struck
(2001) have detected CO in a tail located to the north of SQ at 6000
and 6700 km s$^{-1}$. This emission has H~{\scshape i} and H$\alpha$
counterparts, and the velocities suggest that this gas corresponds to
material removed from NGC 7318B and NGC 7319.

The IRAS data (Yun et al. 1997) show that the far-infrared emission in
SQ is more extended than in other groups and is remarkably similar to
the X-ray emission (see below). This could indicate  the presence of
warm dust in the intragroupo gas.

%**********
\subsection{\small Radio}

Kaftan-Kassim \& Sulentic (1974) observed the radio continuum at 318, 430,
and 606 MHz  and built a map of the emission which shows a connection
between NGC~7331 and the source 4C33.56 across  SQ. However von
Kap-Herr, Haslam, \& Wielebinski
 (1977) observed the same region at 2695 MHz and 
found no evidence of such a connection.

Allen \& Sullivan (1980) observed in the SQ region the 21 cm
H~{\scshape i} at 750 km s$^{-1}$ and 6700 km s$^{-1}$ and concluded
that the low redshift H~{\scshape i} is entirely associated with
NGC~7320, and that the high-redshift H~{\scshape i} is concentrated in
the range 6570--6630 km s$^{-1}$. However, Sulentic \& Arp (1983) have
measured the neutral hydrogen at low velocity ($v \le2200$ km s$^{-1}$)
and found that it is displaced from the optical position of NGC~7320;
they argued that this has been produced by interaction between NGC~7320
and the other members of  SQ.  Peterson \& Shostak (1980) have observed
several positions of the 21 cm line in the inner and outer parts of SQ
and found three extended systems at velocities 5700, 6000, and 6600 km
s$^{-1}$. The galaxies are enveloped by a cloud of H~{\scshape i} with
a size of $\sim$ 100 kpc (Sullivan 1980; Peterson \& Shostak 1980).
Shostak, Sullivan \& Allen (1984) mapped the H~{\scshape i} line in the
region of  SQ with a resolution of 35$^{\prime\prime}$. They built a
map of the gas with resolution of 50 km s$^{-1}$ and confirmed the
existence of three systems at velocities 5700, 6000,  and 6600 km
s$^{-1}$. Most of the emission comes from the outer parts of the
galaxies. The authors argued that the stripping of  gas from the disks
is due to a tidal encounter in the past and a present collision.
Gillespie (1974, 1977) have found several radio sources $\sim$ 15
arcminutes to the north of  SQ.  However, the most probable
interpretation is a group of radio sources at $z\sim$ 0.5 without any
relation with  SQ.

%************
\subsection{\small X-Ray}

X-ray emission in  SQ was detected (Bahcall, Harris, \& Rood 1984) by
the {\it Einstein Observatory}. Other more recent observations have
been obtained with {\it ROSAT} (Sulentic, Pietsch, \& Arp 1995; Pietsch
et al. 1997), and {\it ASCA} (Awaki et al. 1997). Pietsch et al. (1997)
obtained high spatial resolution X-ray images of  SQ with the {\it
ROSAT} HRI detector. This tidal emission amounts to $\sim 5 \times
10^{41}$ erg s$^{-1}$ and is distributed among the following
components: $a$) the galaxies NGC 7319 and NGC 7318A, $b$) the
intergalactic feature between NGC 7319 and NGC 7318B, $c$) a region
situated $\sim 1^{\prime \prime}$ to the north of NGC~7318A,B, and $d$)
 diffuse hot gas (amounting to one third of the total emission).

%************
\subsection{\small Physical picture}

The discrepant redshift of NGC~7320  has received special attention in
the past. In the orthodox view, this is just a chance projection
similar to those found in a large numbers of Hickson compact groups.
However, some authors (Arp 1973; Burbidge \& Burbidge 1961;  Sulentic
\& Arp 1983) have claimed  the existence of a physical connection
between the low- and high-redshift galaxies of SQ.  This claim is based
on statistical arguments or on the numerous signs of distortions found
in optical, IR and radio images of  SQ.  Recently, the {\it Hubble
Space Telescope} has obtained images of  SQ with high resolution (see
http://spaceflightnow.com/news/n0010/26hststephen).  It has been
claimed (Moles 2001) that these images have  definitely solved the
question of the redshift discrepancy of NGC~7320 because it is possible
to resolve individual stars in this galaxy, showing that it is closer
than the rest of the galaxies of the group. However, Arp (2001)
(private comm.), in strong disagreement with this, claims that in these
images it is also possible to resolve individual stars in the
high-redshift members of the group. Furthermore, using the luminosity
of the H~{\scshape ii} region as a distance indicator, Arp (2001)
(private comm.) points out that the low- and high-redshift galaxies of
SQ  should be at the same distance.

The other relevant fact in the discussion on  SQ is the large
difference in velocity ($\sim 1000$ km s$^{-1}$) between NGC~7318B and
the rest of the high-$z$ members of  SQ.  Moles et al. (1998) argue
that NGC~7318B is crossing the group now; as this galaxy has kept its
gas, this current crossing should be occurring for the first time. As a
consequence of this encounter, the interstellar medium of NGC~7318B is
shocking the intragroup medium, especially in the Northern Starburst
Forming (NSF) region (Gallagher et al. 2001)  and in the tidal tail of
NGC~7319. As is expected from numerical simulations and as an
increasing number of observations show, this interaction is producing
intense starbursts in several regions of the group. There are several
features  indicating that this interaction is in fact in progress; for
instance: {\it i\/}) optically, NGC~7318A and NGC~7318B look like a
typical pair merging, {\it ii\/}) there is  radio emission associated
with the interaction between NGC~7319 and NGC~7318B, and {\it iii\/}) a
large number of knots has been observed in the arms of NGC~7318B.
However, Arp (1973) has suggested that with this difference in
velocities the group would dissipate over a very short timescale so
that it would be very unlikely to see them together.

In the standard picture (Moles et al. 1998), NGC~7320C crossed the
group 10$^8$ years ago producing considerable disruptions; for
instance, it removed the H~{\scshape i} in NGC~7319 creating a visible
tail seen today in this galaxy, with part of the gas being deposited in
a region northeast of NGC~7319. The interaction was also probably
responsible for the Seyfert 2-type nuclear activity of NGC~7319. In
this paper, we will show more evidence of the connection between
NGC~7320C and the other galaxies in SQ.

\section{Observations and data reduction}

\begin{table}[htb]
\caption{Summary of the observations} 
\begin{center}
\begin{tabular}{ccccc} 
 $\lambda_{\rm cent}$ (\AA) & FWHM (\AA) & Tel. &$T_{\rm exp}$ (s) \\
\hline
\\
 6611  & 50 & IAC 80     & 11700 \\
 6687  & 50 & IAC 80     & 5400  \\
 6724  & 50 & IAC 80     & 4800 \\
 6569  & 113 & Calar 2.2 m & 1000  \\
 6667  & 76 & Calar 2.2 m & 1000  \\
 6737  & 66 & Calar 2.2 m & 1000  \\
% 6568  & 95 & INT    &       \\
$R$&   &INT    & 2200  \\
$R$& & IAC~80 & 4800\\
\\
\end{tabular}
\end{center}
\end{table}

The observations analyzed in this paper were obtained during several
runs between 1997 and 2000 on the 0.8 m IAC 80 telescope at Teide
Observatory\footnote{The IAC 80 is operated by the Instituto de
Astrof\'\i sica de Canarias at Teide Observatory on the island of
Tenerife (Spain).}, the 2.2 m telescope at Calar Alto\footnote{The
German--Spanish Astronomical Centre, Calar Alto, is operated jointly by
the Max-Planck-Institute for Astronomy, Heidelberg, and Spain's
National Commission for Astronomy.}, and the 2.5 m Isaac Newton
Telescope (INT) at  Roque de los Muchachos Observatory on La Palma
(Spain)\footnote{The INT is operated  by the Isaac Newton Group.}. The
observations taken with the IAC~80 and the 2.2 m telescope, we
concentrated specifically on the H$\alpha$ emission in SQ (the $R$
images taken with the IAC 80 were used only for continuum subtraction
of the H$\alpha$ images). The south of NGC~7331 and SQ were observed in
the $R$-band with the wider field of the WFC on the INT, with a total
exposure time $\sim 2000$ s each. Typical seeing was $\sim $ 2 arcsec
in the narrow band images, and $\sim 1$ arcsec for the $R$-band
observations. Table 1 summarizes the observations.   The spectral
response of the narrow-band filters can be found at
http://\-www.iac.es/\-telescopes/\-ten.html and
http://\-www.caha.es/\-CAHA/\-Instruments/\-filterlist.html. We
performed standard data reduction:  bias subtraction, flat-field
correction, coaddition of images for the same filter, and sky
subtraction. Only the observations at the INT have been photometrically
calibrated, using stars from the catalog of Landolt et al. (1992). For
the observations in the H$\alpha$ narrow bands we subtracted the
continuum using the images in  $R$ taken with the IAC 80 or in a
different redshifted H$\alpha$, scaling according to the flux measured
in field stars.

\section{Broad-band imaging}

Images in the $R$ band were taken with the Wide Field Camera at the 2.5
m INT at Roque de los Muchachos Observatory in 1997  August. The
observations covered $\sim$ 1800 square arcminutes from which we have
selected $\sim 650$ square arcminutes corresponding to the SQ region
and the southern part of the galaxy NGC~7331. We had two objectives:
the first was to delineate the halo of SQ, and the second was to search
for possible evidence of interaction between NGC~7331 and the other SQ
galaxies. The limiting 3$\sigma$ magnitude per 0.37 square arcsecond
pixels was $\sim$ 25.2 mag/arcsec$^2$. However, the structures that we
wanted to map were diffuse and spatially very extended, so we binned
the images to a resolution $\sim$ 2.5 arcsecs, this permits an
enhancement of the signal-to-noise ratio by a factor $\sim$7, that is,
we go almost two magnitudes deeper, being now the 3-$\sigma$ limiting
magnitude 27.3 mag/arcsec$^2$. Figures 1$a,b$ present the gray-scale
and  isophotal map obtained around SQ and in the southern part of
NGC~7331 respectively.  Figure 1$a$ shows that the halo is considerable
more extended that it was revealed in the observations by Moles et al.
(1998). For the first time, we have found evidence of a clear
connection in the form of diffuse light between SQ and the galaxy
NGC~7320C. In particular the SW tail which emerges from SQ seems to
extend  to this galaxy. We think that this supports the previously
suspected connection between SQ and NGC~7320C. We have traced the
emission up to the isophote at 26.7 mag/arcsec$^2$ and 25.5
mag/arcsec$^2$ around the SQ and NGC~7331 respectively. The 25.5
mag/arcsec$^2$ isophote surrounds and seems to enclose NGC~7331. These
numbers are conservative with respect to the sensitivity of the image
and take into account the uncertainties in the estimation of the sky
level, particularly difficult in the observations of the region around
NGC~7331.

There are no signs of past or present interaction in the form of tails,
bridges, etc., between  SQ and NGC~7331. We would remark that the last
isophote surrounding  SQ is not associated with any individual galaxy,
but instead corresponds to the diffuse halo of the group, while the
last isophote detected around NGC~7331 is more regular, roughly
following the shape of the galaxy, and seems to be associated with the
disk (and halo) of this galaxy.

\section{H$\alpha$ maps and velocity field}

To differentiate the emission coming from the low-z (at the systemic
velocity of NGC~7320) and high-z (the other galaxies of SQ)
respectively, we decided to take H$\alpha$ images redshifted at the
velocities of both components (see Section~3). Figure~2 presents the
$R$ band (gray-scale) and the total H$\alpha$ (contours).  The
H$\alpha$ maps presented here are a combination of the emission
associated with the low- and  high-$z$ components. To enhance the
signal with respect to the noise we binned the data to a common pixel
size of 1.06 arcsec; this produces a slight degradation in resolution.
There are many spurious H$\alpha$ features associated with residuals
due to  imperfect continuum subtraction in the stars of the field.  In
general, there is no ambiguity in distinguishing between these
residuals and real H$\alpha$ emission because the latter is
specifically associated with several extended features that we discuss
below. Figure 3 presents the images in the narrow filters observed
combining the H$\alpha$ low-$z$ (contours) and high-$z$ (gray-scale)
filters, respectively. The size of the images is $\sim 6 \times 6$
arcmin$^2$. From these images, it is clear that the low-$z$ emission is
associated with NGC~7320. The high-$z$ emission is concentrated in
three well differentiated structures:  $i$) the center of NGC~7319,
$ii$) an intergroup feature in the  NS direction between NGC~7318B and
NGC~7319, and $iii$) a structure to the north of the aforementioned
feature, which corresponds to features previously identified  by Xu,
Sulentic, \& Tuffs (1999) and Gallagher et al. (2001). Apart from these
components, there are several other high-$z$ H$\alpha$ features in the
tidal tails produced by the strong interaction within the group. From
Fig.~3, we see that the low- and high-$z$ emission are associated with
different spatial regions, and that there are no regions in which both
types of emission overlap. Although our spatial resolution is poor and
the H$\alpha$ emission is diffuse, using the algorithm SExtractor
(Bertin \& Arnouts 1996) and eliminating the spurious detections
associated with star residuals, we identified about 20 and 30 separated
regions in the low- and high-$z$ maps, respectively. The low-$z$
structures are symmetrically distributed in the disk of NGC~7320 (see
however Section~6.1) and tend to be isolated and compact. The high-$z$
emission tends to be more diffuse than the low-$z$ component. This is
in agreement with the standard picture in which  components $ii$ and
$iii$ (see above) of the high-$z$ emission are associated with the
intergroup medium formed by material stripped by recent and ongoing
strong interaction involving the galaxies NGC~7318A, NGC~7318B,
NGC~7319, and NGC~7320C.

We may consider the set of H$\alpha$ images as a two-dimensional
spectrum with very low resolution. It then makes sense to compute the
centroid and dispersion of a given feature. From this, it is
straightforward to produce a two-dimensional velocity map. We proceed
as follows. A single wavelength, $\lambda _i$, was assigned to the
$i$th  pixel of the image:

\begin{equation}
\lambda _i=\frac{\sum _j F_{ij} \times \lambda _{j}}{\sum _j F_{ij}}
\end{equation}

\noindent where $F_{ij}$ the flux in  pixel $i$ and filter $j$, and
$\lambda _{j}$ is the central wavelength in  pixel $i$. To eliminate
most of the residuals due to stars, only fluxes at a level above
2$\sigma$ were considered. From this, we computed the associated
velocity as $v_i=(\lambda_i-\lambda_0)/\lambda_0$ ($\lambda_0=6562$
\AA). Figures~4$a$ and $b$ show the velocity and the associated error
map, respectively. Overplotted in Fig.~4$a$ is the ellipse which fits
the outer isophote of NGC~7320 in the $R$-band. An important result is
that the velocities derived are in two well separated ranges:  one from
5500 to 7500 km s$^{-1}$, and the other from 700 to 1000 km s$^{-1}$.
No features at intermediate velocities that could suggest a Doppler
redshift as the origin of the difference in velocities  between
NGC~7320 and the rest of the SQ have been detected. However, some
points concerning the relationship between these two features are
unexplained and are discussed in next section.

\section{Discussion}

The relation between NGC~7331 and NGC~7320 is clear according to their
relative positions and systemic velocities. From the H~{\scshape i}
line the difference in velocities between both galaxies is $\sim 45$ km
s$^{-1}$ (de Vaucouleurs et al. 1991). Their projected separation is
$\sim 100$ kpc, and the magnitudes in the $B$ band are 9.38 and 13.56
for NGC~7331 and NGC~7320, respectively. These allow us to classify
NGC~7320 as a satellite of NGC~7331 according to the criteria defined
by Zaritsky et al. (1997). We have searched  the NED for objects at
less than 500 kpc from NGC~7331 and with recessional velocities similar
to that of this galaxy and found another two galaxies besides NGC
7320. These galaxies seem to constitute one of the loose groups
identified by Trasarti-Battistoni (1998). At angular distances $\le 1$
degree ($\le 2$ Mpc) from SQ, there are about ten other galaxies with
measured redshifts and with recessional velocities in the range
5500--7500 km s$^{-1}$; they form another loose group with
9 members identified by Trasarti-Battistoni (1998); a subgroup of such
galaxies (NGC7335, NGC7337, and NGC7340) at the East of NGC7331 has
been classified as a poor cluster by White et al. (1999).

One of the mechanisms proposed to explain the counter-rotating core
found in NGC~7331 (Prada et al. 1996) is the interaction of this galaxy
with a small satellite. Although this could really be  the origin of
such kinematics, in our deep $R$ images we have found no evidence
(bridges, tails, etc.)  to $R \sim 26.7$ mag/arcsec$^2$ supporting such
an interaction between NGC~7331 and NGC~7320.

We have carried out a comparative study between the features detected
in our velocity map and  the results by Plana et al. (1999). The
position of these common H$\alpha$ regions is indicated in Fig. 3. We
have indicated also in the figure two features in one of the tails connecting
NGC~7320c with the rest of SQ. The comparison with the
results by Plana et al. (1999) shows that we have detected and
estimated the velocities of the regions detected by these authors (the
detection of region number 23 it is not completely clear in our data),
although we have found a systematic difference of $\sim +300$ km
s$^{-1}$ in the estimated velocities. This is consistent with the
uncertainties expected from the typical width of the H$\alpha$ filters
used in our analysis. In Table~2 we present our estimate of the
velocities of H$\alpha$ regions detected by Plana et al. (1999) (we
have adopted the notation used by these authors). This uncertainty is
too large to distinguish between the components associated with
NGC~7319 and the other three galaxies (NGC~7318A, NGC~7318B, and
NGC~7317), but is small enough to distinguish between emission
associated with the above four galaxies and NGC~7320.  Due to the
technique used, in some cases Plana et al.  (1999) were not able to
measured unambiguously the velocities, so our observations represent
the first determination of such velocities. Our analysis allows us to
distinguish only one component in velocity in each position. If there
were more than one, this should reveal itself in our velocity map as an
increase in the  pixel-to-pixel rms according to the importance of one
or other component. This seems to be the case for region 13 (in the
notation of Plana et al. 1999) in which these authors detected two
different components, and which shows up in our map as a large
variation in velocity from pixel to pixel.

\begin{table}
\caption{Recessional velocities (in km s$^{-1}$) of the H$\alpha$
regions in common with Plana et al. (1999).}
\begin{center}
\begin{tabular}{cccccccc} 
\hline
Region &1 & 2 & 3 & 4 & 5 & 6  \\

Velocity & 6300 & 6000 & 6300 & 6300 & 6300 & 7200   \\
\hline

Region &7 &8 & 9 & 10 & 11 & 12   \\
Velocity & 6900 & 6400 & 6900 & 6300 & 6250 & 6400   \\
\hline

Region & 13 & 14 &15 & 16 & 17 & 18   \\
Velocity & 6100 & 6100 & 5500 & 5950 & 5750 & 6100  \\
\hline
Region & 19 & 20 & 21&22 & 23 \\
Velocity & 6000 & 6200 & 5800 & 5750 & 6100?\\
\end{tabular}
\end{center}
\end{table}

Some of these features are in the tidal tails resulting from the strong
interaction within the group. This is in perfect agreement with the
results of Iglesias-P\'aramo \& V\'\i lchez (2001), who identified these
features as dwarf galaxies. The above results are also in agreement
with the predictions from numerical simulations (Barnes 1988) in which
dwarf galaxies can form in  tidal tails resulting from the
interaction of two disk galaxies.

Our broad-band observations reveal a bigger halo that was previously
observed. As pointed out by Moles et al. (1998), this light is not
associated with the individual galaxies, but is instead  diffuse light
associated with  present and past interactions within the group. In
particular it indicates large transversal motions of NGC~7317 and
NGC~7320C. The interaction with this galaxy seems to be  responsible
for the diffuse light observed to the north of  SQ.  The extension of
the SW tail to NGC~7320C strongly supports the standard scenario
proposed by Moles  et al. (1998),  in which the galaxy crossed  SQ and
created this tail, which runs in parallel to a second tail which
emerges from one of the arms of NGC~7319. This second tail is brighter
as it would be expected if it corresponds to a past encounter with the
core from which most of the gas associated with NGC~7319 (and partially
with NGC~7320C) was stripped.

\subsection{{\bf A remark on some odd coincidences}}

Apart from all the positive facts which suggest that NGC 7320 is
unrelated to the rest of the components of SQ, as we have argued above,
there is one further observational aspect whose interpretation is not
clear in the light of the orthodox view: there is an H$\alpha $ bridge
(see Figs. 3, 4$a$ and 4$c$) which is a continuation of the
intergalactic gas of the high-redshift galaxies of SQ pointing to
NGC~7320 and extending throughout this galaxy. In particular, we would
like to remark:

\begin{itemize}

\item The emission of H$\alpha $ at 6500 km/s is in a region of NGC7320
where there is no emission of H$\alpha $ at 800 km/s. Assuming
elliptical symmetry for NGC 7320 (such symmetry exists in the
$R$-band), we would expect to observe  H$\alpha $ emission with both
velocities (6500 km s$^{-1}$ and 800 km s$^{-1}$) superposed in the
region of the bridge. Instead of that, we observe a gap in the
H${\alpha}$ emission at 800 km/s precisely where the bridge at 6500
km/s is present.

\item Exactly in the point where the bridge of 6500 km/s finishes, HII
regions at 800 km/s begin. We might think that the opacity of  these
HII regions does not allow to see what is behind. However, it is not
only in the position of the HII regions at 800 km/s, but also in the
gaps between them, where we do not find any emission at 6500 km/s.

\end{itemize}

These coincidences could be explained naturally if a bridge at 6500
km/s were in front of the galaxy at 800 km/s. Although the evidence is
not conclusive, and difficult to fit in an ortodox view, we think that
this is a new and interesting observational fact to consider in future
studies of SQ.

\section{Conclusions}

1) We have obtained deep $R$ images in the region from SQ
to the southern edge of the giant spiral NGC~7331. We have found no
evidence of an optical connection in the form of tails, bridges, etc.,
between  SQ and NGC 7331 above magnitudes $\sim 26.7$ mag/arcssec$^2$.

2) We have delineated the halo of SQ up to the above magnitude showing
that it extends to and contains the galaxy NGC~7320C. Additionally, in
the $R$-band image, a tail connecting the core of  SQ and NGC~7320C is
seen. This is interpreted as a tidal tail produced by a past encounter
between NGC~7320C and the core of the group.

3) With narrow-band images in the range 0--8000 km s$^{-1}$, we have
delineated the H$\alpha$ emission. We see that this is concentrated in
a feature in the NS direction situated between NGC~7318B and NGC~7319,
a region $\sim 1$ arcminute to the north of NGC~7318B and the nucleus
of NGC~7319.

4) With the H$\alpha$ images, we have obtained a low-resolution velocity
map of the SQ region. The H$\alpha$ images and the velocity map show
evidence of emission in the tidal tail that is the result of the
strong interaction within the group.

5) The velocity map derived from the H$\alpha$ images does not show
features at intermediate velocities between the low- and the high-$z$
components.

6) A filament in H$\alpha$ at $\sim 6500$ km s$^{-1}$ apparently
connecting NGC~7320 (at $\sim 800$ km s$^{-1}$) and the rest of SQ is
seen. We think that this fact leaves the point of the possible relation
between NGC~7320 and the rest of SQ still open and will require further
attention.

\acknowledgments

We wish to thank  J. Iglesias-P\'aramo, who kindly provided the
H$\alpha$ images taken at the 2.2 m telescope in Calar Alto.

\clearpage

\section*{Figures}

\figcaption[]{Gray-level and contour $R$ map in  the region of  Stephan's Quintet and in the southern part of NGC~7331. The lowest isophotes are 26.7 and 25.5 mag/arcsec$^2$, respectively.}

\figcaption[]{Broad-band $R$  filter (gray-scale) and H$\alpha$ emission (contours) of Stephan's Quintet.}

\figcaption[]{Low-$z$ (contours) and  high-$z$  (gray-scale) H$\alpha$ maps in the region of Stephan's Quintet.}

\figcaption[]{$(a)$ Velocity and $(b)$ error maps in  Stephan's Quintet region obtained from the low-$z$ and high-$z$ H$\alpha$ images. $(c)$ An expanded view of the velocity map centered on the region of abrupt transition from the low to the high-$z$ emission.}

\end{document}